\documentclass[onecolumn,showpacs,superscriptaddress,pre,nofootinbib]{revtex4}

\usepackage{graphicx}
\usepackage{amsmath,amssymb}
\usepackage{color}
\usepackage{amsfonts}

\begin{document}

\newtheorem{theorem}{Theorem}
\newtheorem{definition}{Definition}
\newtheorem{lemma}{Lemma}
\newtheorem{proposition}{Proposition}
\newtheorem{remark}{Remark}
\newtheorem{corollary}{Corollary}
\newtheorem{example}{Example}


\title{Anomalous diffusion on a fractal mesh}

\author{Trifce Sandev}
\email{sandev@pks.mpg.de} \affiliation{Max Planck Institute for the Physics of Complex Systems,
N\"{o}thnitzer Strasse 38, 01187 Dresden, Germany} \affiliation{Radiation Safety
Directorate, Partizanski odredi 143, P.O. Box 22, 1020 Skopje,
Macedonia} \affiliation{Research Center for Computer Science and Information Technologies, Macedonian Academy of Sciences and Arts, Bul. Krste Misirkov 2, 1020 Skopje,
Macedonia}

\author{Alexander Iomin}
\email{iomin@physics.technion.ac.il} \affiliation{Max Planck Institute for the Physics of
Complex Systems, N\"{o}thnitzer Strasse 38, 01187 Dresden, Germany} \affiliation{Department of Physics, Technion, Haifa 32000, Israel}

\author{Holger Kantz}
\email{kantz@pks.mpg.de} \affiliation{Max Planck Institute for the Physics of Complex Systems,
N\"{o}thnitzer Strasse 38, 01187 Dresden, Germany}

\date{\today} 

\begin{abstract}
An exact analytical analysis of anomalous diffusion on a fractal mesh is presented. The fractal mesh structure is a direct product of two fractal sets which belong to a main branch of backbones and side branch of fingers. The fractal sets of both backbones and fingers are constructed on the entire (infinite) $y$ and $x$  axises. To this end we suggested a special algorithm of this special construction. The transport properties of the fractal mesh is studied, in particular, subdiffusion along the backbones is obtained with the dispersion relation $\langle x^2(t)\rangle\sim t^{\beta}$, where the transport exponent $\beta<1$ is determined by the fractal dimensions of both backbone and fingers. Superdiffusion with $\beta>1$ has been observed as well when the environment is controlled by means of a memory kernel.
\end{abstract}

\pacs{87.19.L-, 05.40.Fb, 82.40.-g}
\keywords{fractal mesh, fractal dimension, anomalous diffusion, L\'{e}vy flights}
\maketitle

\section{Introduction}

The transport of particles in inhomogeneous media exhibits anomalous diffusion where the mean square displacement (MSD) behaves with time as $\left\langle x^{2}(t)\right\rangle\simeq t^{\beta}$, where $\beta$ can be \textit{e.g.}, an anomalous transport exponent of fractal structures \cite{RT}. This phenomenon is well established \cite{MS,GAA,MS2,KBS} and well reviewed (see for example, \cite{BG,PSI,AH,sokolov}). A comb model is a simple example of anomalous diffusion \cite{WB,WH} affected by geometry, which however reflects many important transport properties of inhomogeneous media \cite{sokolov,cassi,barkai}, and where due to its specific geometry the MSD bahaves as $t^{1/2}$. It is a particular example of geometrical traps, which can be explained in the framework of the continuous time random walk (CTRW) theory, where the returning probability scales similarly to $t^{-1/2}$, and the waiting times are distributed according $t^{-3/2}$ \cite{MS2,metzler report}. An interesting interplay between the fractal dimensionality and the CTRW in the framework of the comb geometry, suggested in Ref. \cite{fractal grid pre}, leads to increasing the transport exponent $\beta=(1+\nu)/2$ due to the fractal dimension of the backbone $\nu$. This construction leads to a so-called {\it fractal grid comb} model, or {\it fractal grid} \cite{fractal grid pre}. The grid comb model represents a generalization of the comb model, where diffusion along the $x$ direction may occur on many backbones. The corresponding equation for the two dimensional probability distribution function (PDF) is given by the Fokker-Planck equation 
\begin{align}\label{diffusion like eq on a comb two
delta} \frac{\partial}{\partial t}P(x,y,t)
=\mathcal{D}_{x}\sum_{j=1}^{N}w_{j}\delta(y-l_{j})\frac{\partial^{2}}{\partial
x^{2}}P(x,y,t)+\mathcal{D}_{y}\frac{\partial^{2}}{\partial
y^{2}}P(x,y,t),
\end{align}
where $w_{j}$ are structural constants such that $\sum_{j=1}^{N}w_{j}=1$, and the backbones are located at the positions $y=l_{j}$, $j=1,2,\dots,N$, $0\leq
l_{1}<l_{2}<\dots<l_{N}$. The number of backbones $N$ can be arbitrarily large. The initial condition is
\begin{equation}
\label{initial condition}
P(x,y,t=0)=\delta(x)\delta(y),
\end{equation}
and the boundary conditions for the PDF $P(x,y,t)$ and for $\frac{\partial}{\partial q}P(x,y,t)$, $q=\{x,y\}$ are set to zero at infinities, $x=\pm\infty$, $y=\pm\infty$. The case with $l_{1}=0$, $w_{1}=1$, and $w_{2}=w_{3}=\dots=w_{N}=0$ corresponds to the classical comb \cite{havlin}, where anomalous subdiffusion of form $\left\langle x^{2}(t)\right\rangle\simeq t^{1/2}$ takes place. Here $\mathcal{D}_{x}\delta(y)$ is the diffusion coefficient in the $x$ direction with physical dimension $[\mathcal{D}_{x}\delta(y)]=\mathrm{m}^{2}/\mathrm{s}$, $[\mathcal{D}_{x}]=\mathrm{m}^{3}/\mathrm{s}$ ($[\delta(y)]=\mathrm{m}^{-1}$), and $\mathcal{D}_{y}$ is the diffusion coefficient in the $y$ direction with physical dimension $[\mathcal{D}_{y}]=\mathrm{m}^{2}/\mathrm{s}$. For infinite number of backbones with positions belonging to a fractal set $\mathcal{S}_{\nu}$ and the fractal dimension $\nu$ 
($\nu<1$),  one considers the fractal grid comb. 

\begin{figure}\resizebox{0.5\textwidth}{!}{\includegraphics{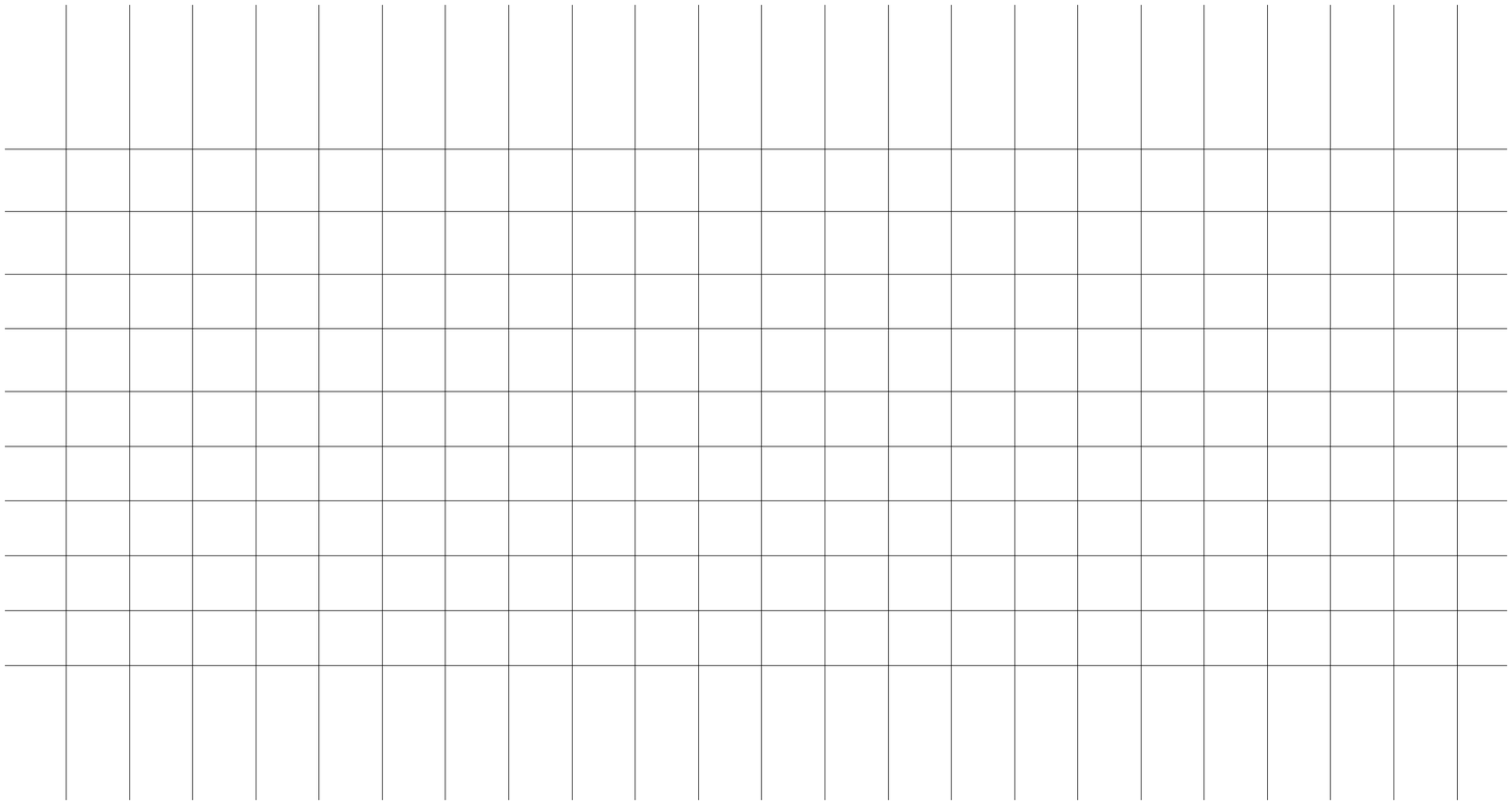}} \caption {Grid with finite number of backbones and continuously distributed fingers along the $x$-axis.}\label{fig_grid}
\end{figure}

In Fig.~\ref{fig_grid} we give an illustration of a grid with a finite number of backbones and continuous distribution of fingers. As it was shown in Ref.~\cite{fractal grid pre}, the finite number of backbones does not change the transport exponent $\beta=1/2$. The situation changes dramatically when the infinite number of backbones ($N\rightarrow\infty$) are distributed inside a fractal volume $l^{\nu}$, then the transport exponent becomes $(1+\nu)/2$. In this paper we consider a general case of fractal distribution of both fingers and backbones, and this construction we call a {\it fractal mesh}.

The main motivation for such a fractal model is its possible application for description of transport properties on porous material, in particular of porous dielectrics, with a low dielectric constant \cite{AKB}. This problem is related to fabrication of porous meta-materials in micro- and nano-electronics \cite{maex}. As it was admitted in Ref. \cite{AKB} ``in an attempt to lower the dielectric constant even more, porosity is being introduced into these new materials". Therefore, the study of diffusion processes in porous dielectrics is of prime importance. 

The paper is organized as follows. In Sec. II we give an explanation of the construction of a fractal set of fingers on infinite axis by mapping a random one third Cantor set, constructed on a circle, onto an infinite line. In Sec. III we treat the case of fractal structure of fingers controlled by the Weierstrass function. Power-law distribution of fingers is considered in Sec. IV. In Sec. V we add a memory kernel which works as an accelerator process for the superdiffusion realization. The summary is given in Sec. VI. Auxiliary material related to the Fox function and Lommel's equation is presented in appendixes.

\section{Fractal set of fingers on infinite axis}

Generalizing the fractal grid model (\ref{diffusion like eq on a comb two delta}), we introduce a two dimensional current $\mathbf{j}=(j_x,\,j_y)$ along the fractal structures of both fingers and backbones, which reads
\begin{eqnarray}\label{JxJy}
\label{current1}&& j_x=-\mathcal{D}_x\sum_{l_{j}\in\mathcal{S}_{\nu}}\delta(y-l_j)\frac{\partial}{\partial x}P(x,y,t),\\
\label{current2}&& j_y=-\mathcal{D}_y\sum_{r_{k}\in\mathcal{S}_{\bar{\nu}}}\delta(x-r_k)\frac{\partial}{\partial y}P(x,y,t).
\end{eqnarray}
The summations in Eqs.~(\ref{current1}) and (\ref{current2}) are over fractal sets $\mathcal{S}_{\nu}$ and $\mathcal{S}_{\bar{\nu}}$ with the fractal dimensions $\nu$ and $\bar{\nu}$, respectively. 

Substituting this current in the Liouville equation,
\begin{align}
\label{liouville eq}
\frac{\partial}{\partial t}P+\mathrm{div}\,\mathbf{j}=0,
\end{align}
one obtains
\begin{align}\label{fractal mesh} 
\frac{\partial}{\partial t}P(x,y,t)
=\mathcal{D}_{x}\sum_{l_j\in\mathcal{S}_{\nu}}\delta(y-l_{j})\frac{\partial^{2}}{\partial
x^{2}}P(x,y,t)+\mathcal{D}_{y}\sum_{r_k\in\mathcal{S}_{\bar{\nu}}}\delta(x-r_{k})\frac{\partial^{2}}{\partial y^{2}}P(x,y,t).
\end{align}
This model represents a two dimensional grid structure, or \textit{a fractal mesh} with infinite number of backbones and fingers at positions which belong to fractal sets $\mathcal{S}_{\nu}$ and $\mathcal{S}_{\bar{\nu}}$, respectively. Here we note that $\mathcal{D}_{x}$ and $\mathcal{D}_{y}$ are generalized diffusion coefficients with physical dimensions $\left[\mathcal{D}_{x}\right]=\mathrm{m}^{2-\nu}/\mathrm{s}$ and $\left[\mathcal{D}_{y}\right]=\mathrm{m}^{2-\bar{\nu}}/\mathrm{s}$ that absorb the dimension of fractal volumes $l^{\nu}$ and $l^{\bar{\nu}}$, respectively \cite{fractal grid pre}.  

\begin{figure}\resizebox{0.5\textwidth}{!}{\includegraphics{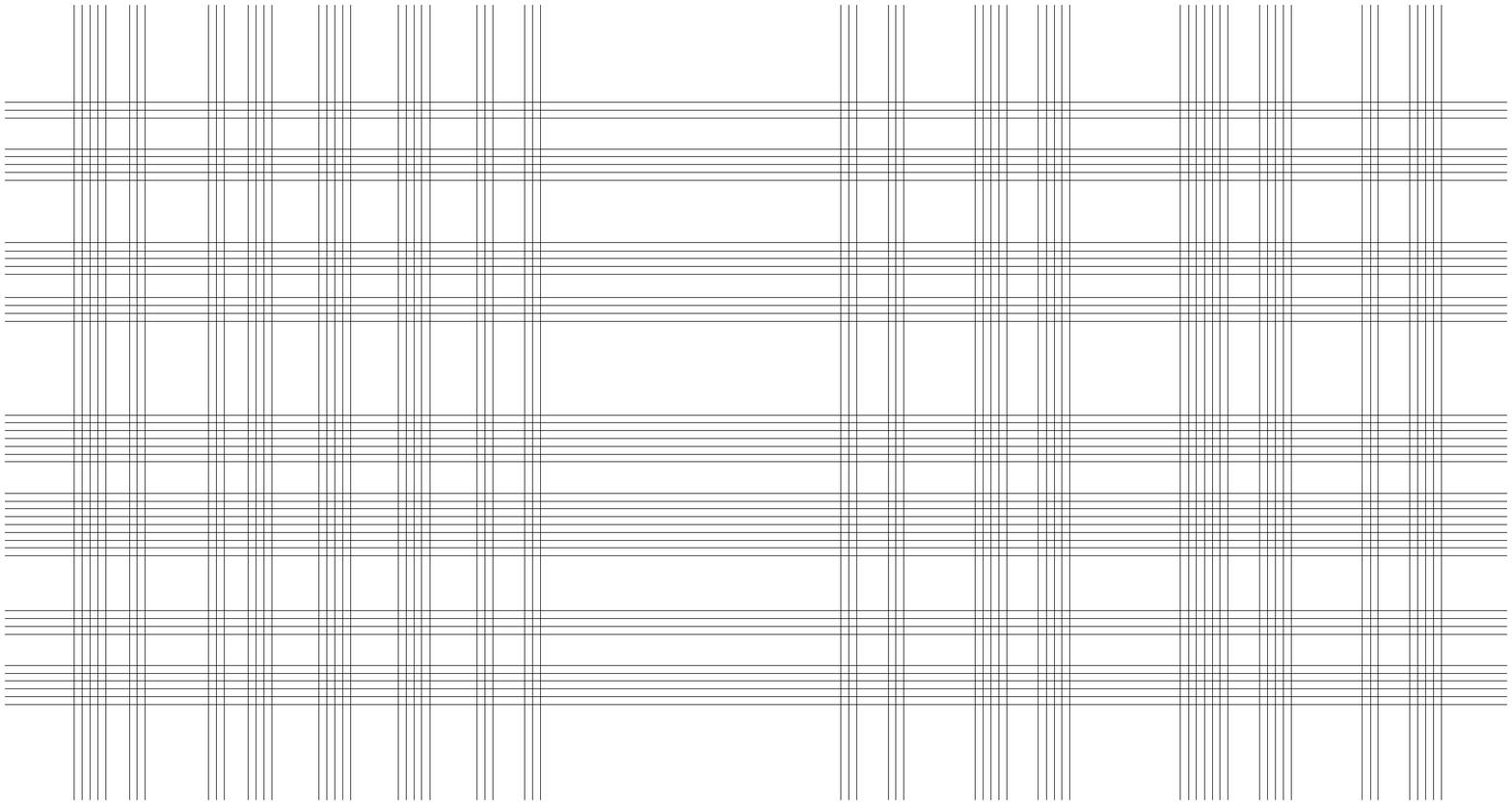}} \caption {A fractal mesh, where random one third Cantor set of backbones (third generation of construction) can be distributed on either a finite strip, or entire $y$ axis, while the  fractal set of fingers is placed on the entire $x$ axis with fifth generation of construction.}\label{fig2}
\end{figure}

Illustrations of fractal mesh construction is plotted in Fig. \ref{fig2}, where the fractal structure of fingers/backbones is a random form of a middle third Cantor set \cite{falconer}. The algorithm of the construction is as follows. A given segment is randomly divided in three parts and the middle part is removed. Therefore, the first generation consists of two subsets. This middle third procedure is repeated for each subset to obtain the second generation with four random subsets of continuously distributed fingers, or backbones. Then, one obtains the third generation with eight random subsets, etc. In Fig. \ref{fig2}, we consider a fractal structure of both backbones and fingers, which corresponds to the fractal mesh structure.

\begin{figure}\resizebox{0.5\textwidth}{!}{\includegraphics{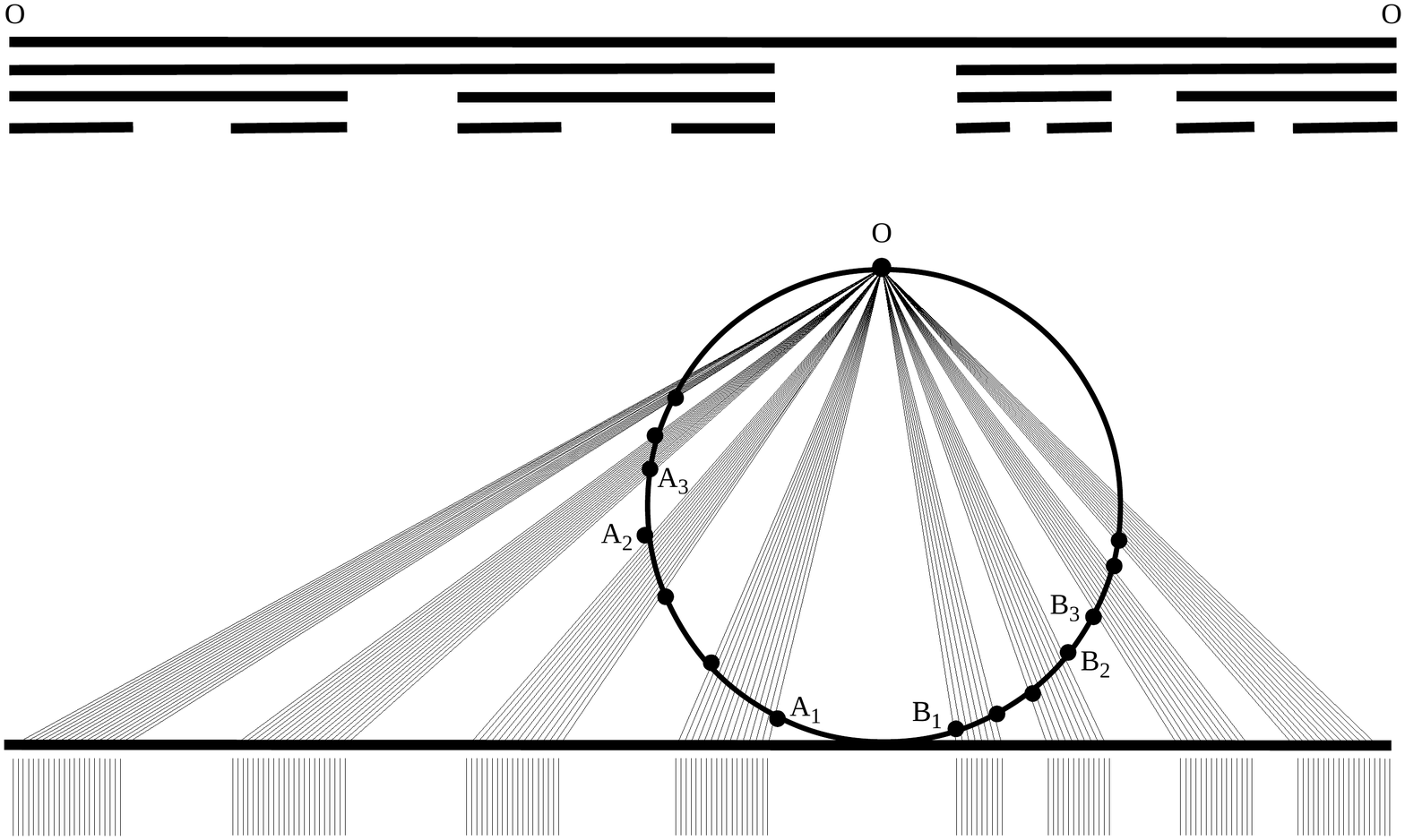}} \caption {Construction of fractal grid of fingers on the infinite axis. The fractal set on the infinite axis is constructed by projecting the fractal set on a circle from point O, which belongs to the fractal set as well. Its projection on the $x$ axis corresponds to $\pm \infty$, which ensures the boundary condition on infinities.}\label{fig_circle}
\end{figure}

While construction of the fractal set on the finite segment $[-L,L]$ is straightforward, construction of the fractal set on the infinite axis needs some care. Therefore, we suggest an algorithm of construction of random third middle Cantor set on the infinite $x$ axis. This algorithm of mapping of a random third middle Cantor set is only illustrative, however it should work for any fractal set. To this end we take into account that the power of the finite segment $[0,1]$ is the same as the power of an infinite line. Fig.~\ref{fig_circle} illustrates this algorithm, where we present the unit segment in the form of a circle by closing the end points (end points O on the top of the circle). Then we follow the previous procedure of a random one third Cantor set construction. We randomly divide the circle in three parts by points $\mathrm{O}$, $\mathrm{A}_1$ and $\mathrm{B}_1$, and remove central segment $\mathrm{A}_{1}\mathrm{B}_{1}$. From point $\mathrm{O}$ we draw rays passing through $\mathrm{A}_1$ and $\mathrm{B}_1$, and intersect the horizontal line. Next, we divide each segment $\mathrm{OA}_1$ and $\mathrm{OB}_{1}$ randomly each in three parts by points $\mathrm{A}_2$ and $\mathrm{A}_3$, and $\mathrm{B}_2$ and $\mathrm{B}_3$, respectively, and remove central segments $\mathrm{A}_2\mathrm{A}_3$ and $\mathrm{B}_2\mathrm{B}_3$. In the same way, we draw rays from $\mathrm{O}$ passing through $\mathrm{A}_2$, $\mathrm{B}_2$, $\mathrm{A}_3$ and $\mathrm{B}_3$, which intersect the horizontal line. We follow this procedure of random division of the segments $\mathrm{A}_{1}\mathrm{A}_{2}$, $\mathrm{B}_{1}\mathrm{B}_{2}$, $\mathrm{A}_{3}\mathrm{O}$, $\mathrm{B}_{3}\mathrm{O}$ in three parts, and we remove the middle segment. Then, we map the random middle third Cantor set of points on the unit circle onto the infinite axis by directing rays from O through the each point of the circle Cantor set to the $x$ axis, producing a fractal set on the infinite real axis with the same fractal dimension \cite{milovanov}. Considering fingers on the horizontal line we construct a fractal set of fingers on the infinite $x$ axis. It is worth noting that the ends O of the unit segment, belonging to the circle fractal set, also belong to the fractal set of the $x$ axis  since the intersections of the rays from O with the horizontal line are at infinities. Therefore, this projection also ensures the existence of the boundary conditions at infinities.

\section{Fractal mesh model}\label{Sec4}

Fractals can be described by means of the Weierstrass function \cite{berry,west}. We apply this function to construct the fractal set of the fingers. We rewrite the last term of Eq.~(\ref{fractal mesh}), in the form 
\begin{align}\label{last term eq}
\mathcal{D}_{y}\frac{\partial^{2}}{\partial
y^{2}}\sum_{k=1}^{\infty}w_{k}\,\delta(x-r_{k})P(x,y,t).
\end{align}
It describes diffusion along the fingers located at $x=r_{k}$, $k=1,2,...$, which belong to the fractal set $S_{\bar{\nu}}$ with fractal dimension $0<\bar{\nu}<1$. Taking the structural constant in the form 
\begin{align}\label{strucctural constants}
w_{k}=\frac{l-b}{b}\left(\frac{b}{l}\right)^{k},
\end{align}
where $l,b>1$, $l-b\ll b$ ($l$ and $b$ are dimensionless scale parameters), we obtain that 
$\sum_{k=1}^{\infty}w_{k}=\frac{l-b}{l}\sum_{k=0}^{\infty}\left(\frac{b}{l}\right)^{k}=1$. In the Fourier $\left(\kappa_{x}\right)$ space, the last term of Eq.~(\ref{fractal mesh}) (given by Eq.~(\ref{last term eq})) reads
\begin{eqnarray}\label{eq in kx space}
\mathcal{D}_{y}\frac{\partial^{2}}{\partial
y^{2}}\sum_{k=1}^{\infty}w_{k}e^{i\kappa_{x}r_{k}}P(x=r_{k},y,t)&&=\mathcal{D}_{y}\frac{\partial^{2}}{\partial
y^{2}}\sum_{k=1}^{\infty}w_{k}e^{i\kappa_{x}r_{k}}\frac{1}{2\pi}\int_{-\infty}^{\infty}d\kappa_{x}{'}P(\kappa_{x}{'},y,t)e^{-i\kappa_{x}{'}r_{k}}\nonumber\\&&=\mathcal{D}_{y}\frac{\partial^{2}}{\partial
y^{2}}\frac{1}{2\pi}\int_{-\infty}^{\infty}d\kappa_{x}{'}\Psi\left(\kappa_{x}-\kappa_{x}{'}\right)P(\kappa_{x'},y,t),
\end{eqnarray}
Here we introduce a so called Weierstrass function (see \textit{e.g.}, \cite{west})
\begin{equation}\label{Weierstrass}
\Psi(z)=\frac{l-b}{b}\sum_{k=1}^{\infty}\left(\frac{b}{l}\right)^{k}\exp\left(i\frac{z}{l^{k}}\right),
\end{equation}
where $r_{k}=L/l^{k}$, $z=\left(\kappa_{x}-\kappa_{x}{'}\right)L$, and $L=1$. One obtains it in the form of scaling function
\begin{equation}\label{Weierstrass2}
\Psi(z/l)=\frac{l}{b}\Psi(z)-\frac{l-b}{b}\exp\left(i\frac{z}{l}\right),
\end{equation}
and by neglecting the last term ($l-b\ll b$), we arrive to the following scaling
\begin{equation}\label{Weierstrass3}
\Psi(z/l)\simeq\frac{l}{b}\Psi(z).
\end{equation}
This scaling satisfies the Weierstrass function  of the power law form $\Psi(z)\sim\frac{1}{z^{1-\bar{\nu}}}$, where
$\bar{\nu}=\log{b}/\log{l}$, $0<\bar{\nu}<1$, is the fractal dimension \cite{berry,west}. Eventually, relation (\ref{eq in kx space}) can be obtained in the convolution form
\begin{align}\label{eq in FF space2}
\mathcal{D}_{y}\frac{\partial^{2}}{\partial y^{2}}\frac{1}{2\pi}\int_{-\infty}^{\infty}d\kappa{'}_{x}\,\Psi\left(\kappa_{x}-\kappa{'}_{x}\right)P(\kappa{'}_{x},y,t)=\frac{\mathcal{D}_{y}}{2\pi}\frac{\partial^{2}}{\partial y^{2}}\int_{-\infty}^{\infty}d\kappa_{x}{'}\,\frac{\tilde{P}(\kappa_{x}{'},y,t)}{|\kappa_{x}-\kappa_{x}{'}|^{1-\bar{\nu}}},
\end{align}
which is the Riesz fractional integral \cite{SKM book} in the reciprocal Fourier space\footnote{The Fourier transform of $f(x)$ is given by $f(\kappa)=\mathcal{F}\left[f(x)\right]=\int_{-\infty}^{\infty}{d}x\,f(x)e^{\imath\kappa x}$. Therefore, the inverse Fourier transform is defined by $f(x)=\mathcal{F}^{-1}\left[f(\kappa)\right]=\frac{1}{2\pi}\int_{-\infty}^{\infty}{d}\kappa\,f(\kappa)e^{-\imath\kappa x}$.}. It is worth noting the specific property of this construction of the fractal set of fingers. When $\bar{\nu}=0$ the fractal dimension of fingers is one. Therefore, the fractal dimension of fingers in real space is $1-\bar{\nu}$.

Performing the Fourier inversion, we arrive at the fractal mesh equation
\begin{align}\label{fractal mesh weierstrass} 
\frac{\partial}{\partial t}P(x,y,t)
=\mathcal{D}_{x}\sum_{l_j\in\mathcal{S}_{\nu}}\delta(y-l_{j})\frac{\partial^{2}}{\partial
x^{2}}P(x,y,t)+\mathcal{D}_{y}C_{\bar{\nu}}|x|^{-\bar{\nu}}\frac{\partial^{2}}{\partial y^{2}}P(x,y,t),
\end{align}
where $C_{\bar{\nu}}=\Gamma(\bar{\nu})\cos{\frac{\bar{\nu}\pi}{2}}$. By the Laplace transform\footnote{The Laplace transform of a given function $f(t)$ is defined by $f(s)=\mathcal{L}[f(t)]=\int_{0}^{\infty}dt\,e^{-st}f(t)$.}, one finds
\begin{align}\label{fractal mesh weierstrass L} 
sP(x,y,s)-P(x,y,t=0)
=\mathcal{D}_{x}\sum_{l_j\in\mathcal{S}_{\nu}}\delta(y-l_{j})\frac{\partial^{2}}{\partial
x^{2}}P(x,y,s)+\mathcal{D}_{y}C_{\bar{\nu}}|x|^{-\bar{\nu}}\frac{\partial^{2}}{\partial y^{2}}P(x,y,s).
\end{align}
We look for the solution of the Eq.~(\ref{fractal mesh weierstrass L}) in the form of the ansatz
\begin{align}\label{ansatz}
P(x,y,s)=g(x,s)\exp\left(-\sqrt{\frac{s}{\mathcal{D}_{y}C_{\bar{\nu}}}}|x|^{\bar{\nu}/2}|y|\right).
\end{align}

From here we calculate the reduced PDF along the backbones $p_{1}(x,t)=\int_{-\infty}^{\infty}dy\,P(x,y,t)$. Therefore, in the Laplace space, we find
\begin{align}
p_{1}(x,s)=2g(x,s)\sqrt{\frac{\mathcal{D}_{y}C_{\bar{\nu}}}{s}}|x|^{-\bar{\nu}/2}.
\end{align}
Integrating Eq.~(\ref{fractal mesh weierstrass}) over $y$, one obtains
\begin{align}\label{fractal mesh weierstrass L p1} 
sp_{1}(x,s)-p_{1}(x,t=0)
=\mathcal{D}_{x}\eta(s)\frac{\partial^{2}}{\partial
x^{2}}\sum_{l_j\in\mathcal{S}_{\nu}}P(x,y=l_j,s),
\end{align}
where the initial condition $p_1(x,t=0)=\delta(x)$ reads from Eq.~(\ref{initial condition}). From Eq.~(\ref{ansatz}) it follows
\begin{align}\label{ansatz2}
P(x,y=l_{j},s)=g(x,s)\exp\left(-\sqrt{\frac{s}{\mathcal{D}_{y}C_{\bar{\nu}}}}|x|^{\bar{\nu}/2}|l_{j}|\right).
\end{align}
The summation in Eq.~(\ref{fractal mesh weierstrass}) is over the fractal set $\mathcal{S}_{\nu}$, which corresponds to integration over the fractal measure $\mu_{\nu}\sim l^{\nu}$, such that $\sum_{l_{j}\in\mathcal{S}_{\nu}}\delta(l-l_{j})\rightarrow\frac{1}{\Gamma(\nu)}t^{\nu-1}$ is the fractal density, and $d\mu_{\nu}=\frac{1}{\Gamma(\nu)}l^{\nu-1}dl$ \cite{tarasov}. Therefore, one finds
\begin{align}\label{sum fractal mesh weierstrass p1} 
\sum_{l_j\in\mathcal{S}_{\nu}}P(x,y=l_j,s)&=g(x,s)\frac{1}{\Gamma(\nu)}\int_{0}^{\infty}dl\,l^{\nu-1}\exp\left(-\sqrt{\frac{s}{\mathcal{D}_{y}C_{\bar{\nu}}}}|x|^{\bar{\nu}/2}l\right)\nonumber\\&=g(x,s)\left(\frac{\mathcal{D}_{y}C_{\bar{\nu}}}{s|x|^{\bar{\nu}}}\right)^{\nu/2}=\frac{s^{(1-\nu)/2}|x|^{\bar{\nu}(1-\nu)/2}}{2\left(\mathcal{D}_{y}C_{\bar{\nu}}\right)^{(1-\nu)/2}}p_{1}(x,s),
\end{align}
where the finite result is obtained by means of Eq.~(\ref{ansatz}). It is worth noting that this integration over the infinite scale is well defined, since the algorithm  of the fractal set construction is well defined as well. However, the obtained result can be also obtained for the finite fractal set of backbones embedded inside a finite segment $[-L\, ,L]$, such that integration in Eq.~(\ref{sum fractal mesh weierstrass p1}) is performed  from $0$ to $L$, which leads to the incomplete gamma function \cite{fractal grid pre}. Eventually taking the limit $L\rightarrow\infty$, we obtain the result in Eq. (\ref{sum fractal mesh weierstrass p1}). Substituting result (\ref{sum fractal mesh weierstrass p1}) in Eq.~(\ref{fractal mesh weierstrass L p1}), one obtains
\begin{align}\label{fractal mesh weierstrass L p1 final} 
s^{-(1-\nu)/2}\left[sp_{1}(x,s)-\delta(x)\right]
=\frac{\mathcal{D}_{x}}{2\left(\mathcal{D}_{y}C_{\bar{\nu}}\right)^{(1-\nu)/2}}\frac{\partial^{2}}{\partial
x^{2}}\left(|x|^{\bar{\nu}(1-\nu)/2}p_{1}(x,s)\right).
\end{align}
After the substitution $f(x,s)=|x|^{\bar{\nu}(1-\nu)/2}p_{1}(x,s)$, Eq.~(\ref{fractal mesh weierstrass L p1 final}) becomes
\begin{align}\label{fractal mesh weierstrass L f final} 
s|x|^{-\bar{\nu}(1-\nu)/2}f(x,s)-\frac{\mathcal{D}_{x}}{2\left(\mathcal{D}_{y}C_{\bar{\nu}}\right)^{(1-\nu)/2}}s^{(1-\nu)/2}\frac{\partial^{2}}{\partial
x^{2}}f(x,s)=\delta(x),
\end{align}
which can be solved following the procedure suggested in \cite{jpa2016}. First we consider the homogeneous part of the equation, which reads
\begin{align}\label{fractal mesh weierstrass L f final homogeneous} 
\frac{2\left(\mathcal{D}_{y}C_{\bar{\nu}}\right)^{(1-\nu)/2}}{\mathcal{D}_{x}}s^{(1+\nu)/2}|x|^{-\bar{\nu}(1-\nu)/2}G(x,s)=\frac{\partial^{2}}{\partial x^{2}}G(x,s).
\end{align}
Eq.~(\ref{fractal mesh weierstrass L f final homogeneous}) is the Lommel equation, and can be solved exactly (see Appendix \ref{app 2}), with the solution
\begin{align}\label{f(x,s) green sol K delta}
G(x,s)=\sqrt{x}K_{\frac{1}{\alpha}}\left(\frac{2s^{\mu/2}x^{\alpha/2}}{\alpha}\sqrt{\frac{2\left(\mathcal{D}_{y}C_{\bar{\nu}}\right)^{1-\mu}}{\mathcal{D}_{x}}}\right)=\frac{\sqrt{x}}{2}H_{0,2}^{2,0}\left[\left.\frac{s^{\mu}x^{\alpha}}{\alpha^{2}}\frac{2\left(\mathcal{D}_{y}C_{\bar{\nu}}\right)^{1-\mu}}{\mathcal{D}_{x}}\right|\left.\begin{array}{l} \\
(\frac{1}{2\alpha},1),(-\frac{1}{2\alpha},1)\end{array}\right.\right],
\end{align} 
where $\alpha=\frac{4-\bar{\nu}(1-\nu)}{2}$, $\mu=(1+\nu)/2$, $K_{1/\alpha}(z)$ is the modified Bessel function (of the third kind) and $H_{p,q}^{m,n}(z)$ is the Fox $H$-function (see Appendices \ref{app solution} and \ref{app 2}).

\begin{figure}\resizebox{0.5\textwidth}{!}{\includegraphics{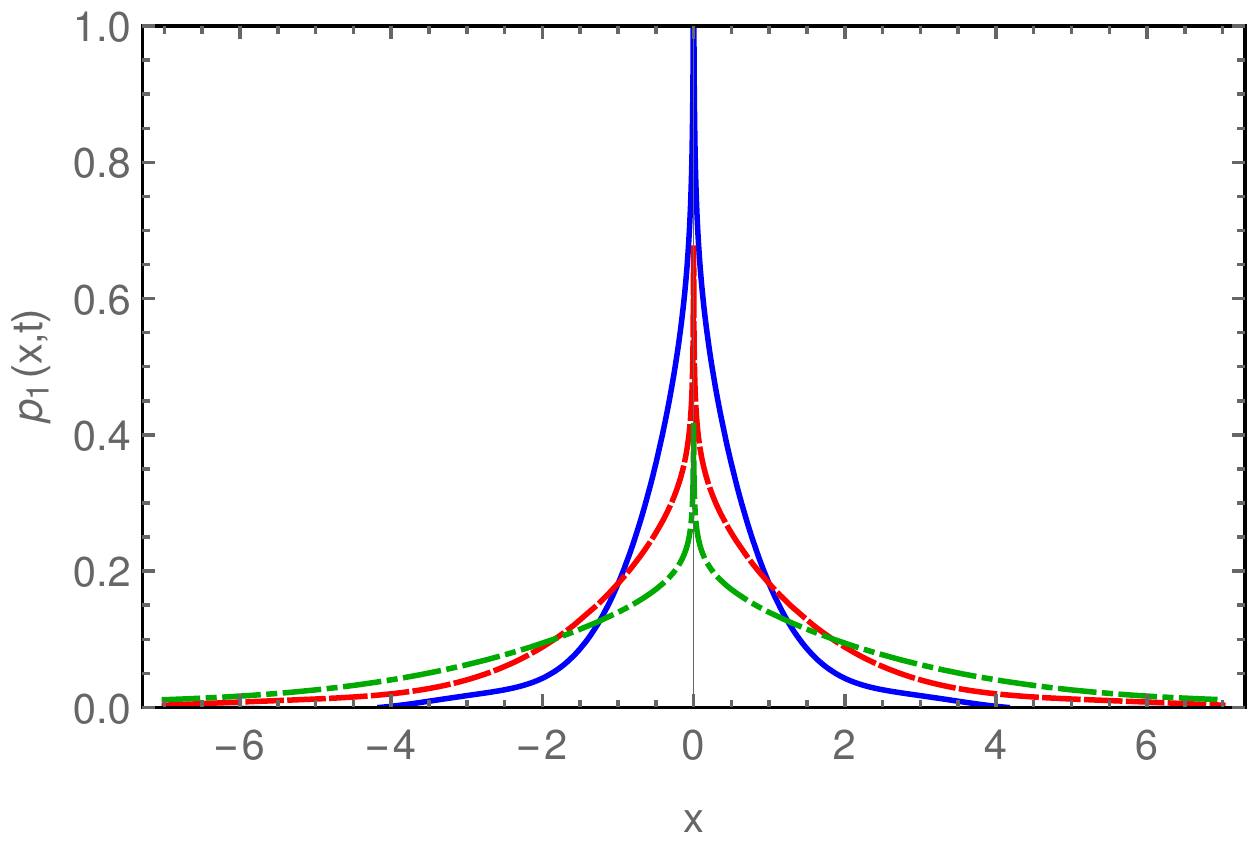}} \caption {Graphical representation of the PDF (\ref{p1(x,t) final exact}) for $\nu=\bar{\nu}=1/2$, $\mathcal{D}_{x}=1$, $\mathcal{D}_{y}=1$, and $t=1$ (solid blue line), $t=5$ (red dashed line), $t=20$ (green dot-dashed line).}\label{pdf mesh}
\end{figure}

Considering the inhomogeneous Lommel Eq.~(\ref{fractal mesh weierstrass L f final}), we use the function $f(|x|,s)=\mathcal{C}(s)G(|x|,s)=\mathcal{C}(s)G(y,s)$, where $G(y,s)$ is obtained in Eq.~(\ref{f(x,s) green sol K delta}), $y=|x|$, and $\mathcal{C}(s)$ is a function which depends on $s$. Thus, we find 
\begin{align}\label{condition}
-2\left[\frac{\mathcal{D}_{x}}{2\left(\mathcal{D}_{y}C_{\bar{\nu}}\right)^{1-\mu}}s^{1-\mu}\right]\frac{\partial}{\partial y}f(y=0,s)=1.
\end{align}
From this equation, by using series representation of the modified Bessel function (\ref{K series}), we find $\mathcal{C}(s)$, and by inverse Laplace transform (see relation (\ref{H_laplace}) in Appendix \ref{app 2}) we finally obtain 
\begin{align}\label{p1(x,t) final exact}
p_{1}(x,t)=\frac{\left(\frac{2\left(\mathcal{D}_{y}C_{\bar{\nu}}\right)^{1-\mu}}{\mathcal{D}_{x}}\right)^{1-\frac{1}{2\alpha}}}{2\alpha^{1-1/\alpha}\Gamma\left(1-1/\alpha\right)}\frac{|x|^{\alpha-3/2}}{t^{\left(1-\frac{1}{2\alpha}\right)\mu}} H_{1,2}^{2,0}\left[\left.\frac{1}{\alpha^{2}}\frac{2\left(\mathcal{D}_{y}C_{\bar{\nu}}\right)^{1-\mu}}{\mathcal{D}_{x}}\frac{x^{\alpha}}{t^{\mu}}\right|\left.\begin{array}{l} (1-\mu+\frac{\mu}{2\alpha},\mu)\\
(\frac{1}{2\alpha},1),(-\frac{1}{2\alpha},1)\end{array}\right.\right].
\end{align}
Graphical representation of solution (\ref{p1(x,t) final exact}) is plotted in Fig.~\ref{pdf mesh}.

\begin{figure}\resizebox{0.5\textwidth}{!}{\includegraphics{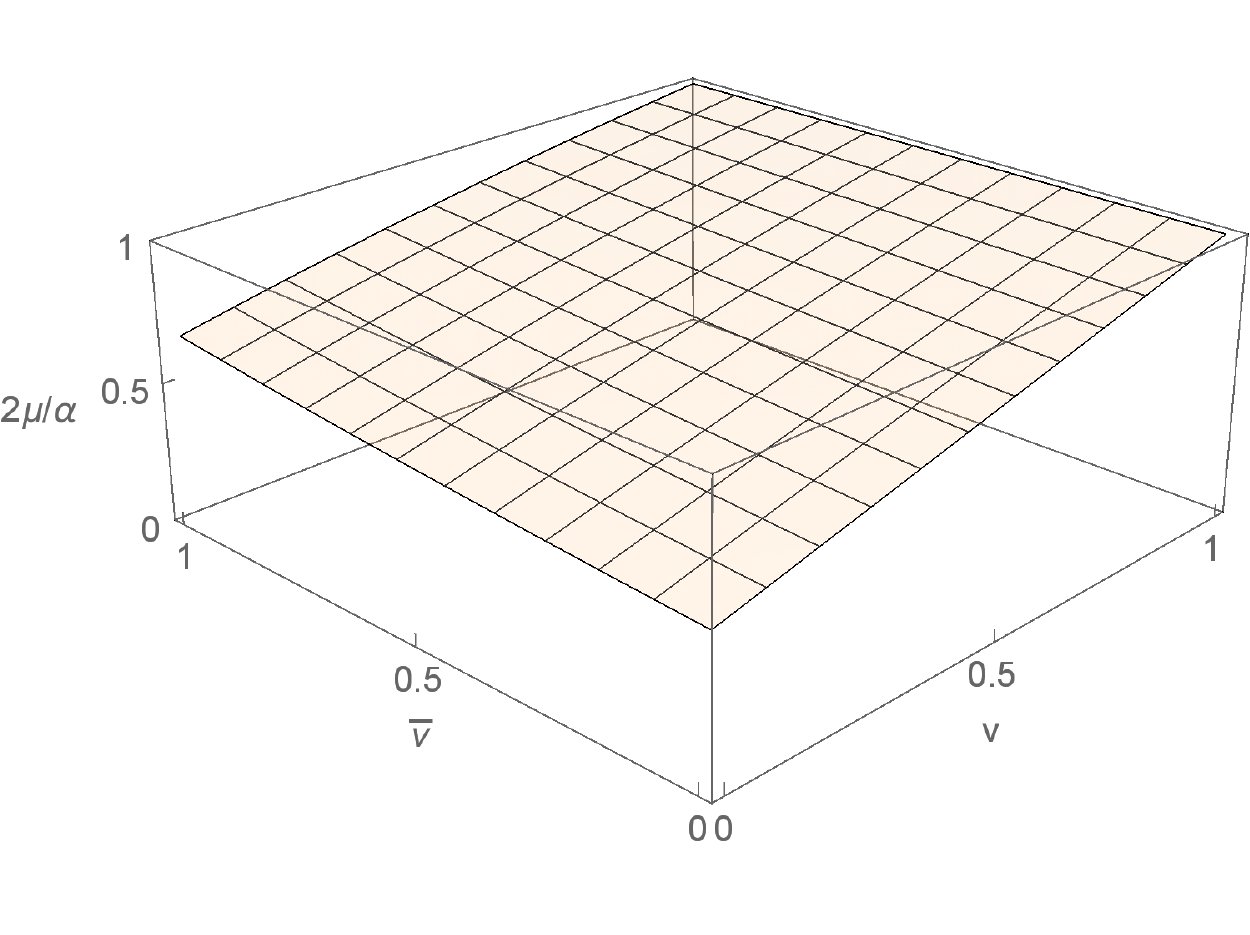}} \caption {The transport exponent $2\mu/\alpha$ vs $\nu$ and $\bar{\nu}$.}\label{fig exponent w}
\end{figure}

From here we calculate the MSD, which reads 
\begin{align}\label{msd w}
\left\langle x^{2}(t)\right\rangle\simeq t^{2\mu/\alpha}=t^{\frac{2(1+\nu)}{4-\bar{\nu}(1-\nu)}}.
\end{align}
This corresponds to subdiffusion with the transport exponent $\frac{1}{2}<\frac{2(1+\nu)}{4-\bar{\nu}(1-\nu)}<1$. Since the latter is larger than $1/2$,
one observes enhanced subdiffusion in comparison to the classical comb where the transport exponent is equal to $1/2$. This results from the fractal structure of the fingers and backbones. Graphical representation of the transport exponent $2\mu/\alpha$ is given in Fig.~\ref{fig exponent w}. It follows that for $\nu=\bar{\nu}=0$, the transport exponent is equal to $1/2$. For $\nu=0$, which corresponds to diffusion along the one backbone only, the transport exponent is $\frac{2}{4-\bar{\nu}}$ \cite{jpa2016} and increases from $1/2$ to $2/3$ by increasing $\bar{\nu}$. For $\bar{\nu}=0$, which means a continuous distribution of fingers, the transport exponent is $\frac{1+\nu}{2}$, and it increases from $1/2$ to $1$ by increasing $\nu$. For $\nu=1$ (particle moves in a finite strip along the $x$ axis) the transport exponent is equal to one for any value of $\bar{\nu}$, which corresponds to the two dimensional normal diffusion.

\subsection{Stationary solution}

The power-law scaling of the backbones can be also described by the Weierstrass function in the same way as the power-law scaling of fingers. In this case the fractional dynamics on the fractal mesh is described by the inhomogeneous diffusion coefficients 
$\mathcal{D}_{x}\mathcal{C}_{\nu}|y|^{-\nu}$ and $\mathcal{D}_{y}\mathcal{C}_{\bar{\nu}}|x|^{-\bar{\nu}}$, and the Fokker-Planck Eqs.~(\ref{fractal mesh}) and (\ref{fractal mesh weierstrass}) now become
\begin{align}\label{fractal mesh weierstrass new} 
\frac{\partial}{\partial t}P(x,y,t)
=\mathcal{D}_{x}\mathcal{C}_{\nu}|y|^{-\nu}\frac{\partial^{2}}{\partial x^{2}}P(x,y,t) +\mathcal{D}_{y}\mathcal{C}_{\bar{\nu}}|x|^{-\bar{\nu}}\frac{\partial^{2}}{\partial y^{2}}P(x,y,t).
\end{align}
Contrary to Eq.~(\ref{fractal mesh weierstrass}), Eq.~(\ref{fractal mesh weierstrass new}) is not integrable due to the complicated interaction between $(x,y)$ degrees of freedom. However, one can study the result of relaxation at infinite times in the framework of the stationary solutions of Eq.~(\ref{fractal mesh weierstrass new}). It is symmetric in respect to $x\rightarrow-x$ and $y\rightarrow-y$, therefore we analyze it on the positive part of the $2D$ plain and then extend the obtained solution symmetrically on the entire plain. Therefore, the equation can be simplified
\begin{align}\label{fractal mesh weierstrass2} 
\frac{\partial}{\partial t}P(x,y,t)
=\mathcal{D}_{x}\mathcal{C}_{\nu}y^{-\nu}\frac{\partial^{2}}{\partial x^{2}}P(x,y,t) +\mathcal{D}_{y}\mathcal{C}_{\bar{\nu}}x^{-\bar{\nu}}\frac{\partial^{2}}{\partial y^{2}}P(x,y,t).
\end{align}
Considering the stationary solution $P_{\mathrm{st}}(x,y)$ of the equation, $\frac{\partial}{\partial t}P_{\mathrm{st}}=0$, we use the separation $P_{\mathrm{st}}(x,y)=X(x)Y(y)$, which yields
\begin{align}\label{fractal mesh weierstrass221} 
0=\mathcal{D}_{x}\mathcal{C}_{\nu}y^{-\nu}\frac{\partial^{2}}{\partial x^{2}}X(x)Y(y) +\mathcal{D}_{y}\mathcal{C}_{\bar{\nu}}x^{-\bar{\nu}}\frac{\partial^{2}}{\partial y^{2}}X(x)Y(y),
\end{align}
or equivalently
\begin{align}\label{fractal mesh weierstrass222} 
\frac{\mathcal{D}_{x}}{\mathcal{C}_{\bar{\nu}}}x^{\bar{\nu}}\frac{X''(x)}{X(x)}=-\frac{\mathcal{D}_{y}}{\mathcal{C}_{\nu}}y^{\bar{\nu}}\frac{Y''(y)}{Y(y)}=\lambda,
\end{align}
where $\lambda$ is a separation constant, or the spectrum, which is determined from the boundary conditions. From here we have the Lommel equations (see (\ref{Lommel}))
\begin{align}\label{fractal mesh weierstrass223} 
X''(x)-\lambda\frac{\mathcal{C}_{\bar{\nu}}}{\mathcal{D}_{x}}x^{-\bar{\nu}}X(x)=0,
\end{align}
\begin{align}\label{fractal mesh weierstrass224} 
Y''(y)+\lambda\frac{\mathcal{C}_{\nu}}{\mathcal{D}_{y}}y^{-\nu}Y(y)=0.
\end{align}
The solutions of these equations are given in terms of the Bessel function $Z_{\xi}(z)$ (see Appendix \ref{app 2}).
However, the solution of Eq.~(\ref{fractal mesh weierstrass224}) does not corresponds to the zero boundary conditions. Therefore, as it is anticipated, there is no any nontrivial stationary solution.

\section{Fractal grid comb}\label{sec grid comb}

A mathematical presentation of the fractal distribution of fingers in the fractal mesh model (\ref{fractal mesh}) can have a various realizations. In particular, it can be a realization of a convolution integral in real space \cite{iomin2}. It is convenient to present it by means of the inverse Fourier transform.\footnote{The density of fingers is $\int dx\,\rho(x)$, where $\rho(x)$ is given by $\mathcal{F}\left[\rho(x)\right]=|\kappa_{x}|^{1-\bar{\nu}}$.}. The correspondingly modified fractal mesh model reads 
\begin{align}\label{fractal grid comb} 
\frac{\partial}{\partial t}P(x,y,t)
=\mathcal{D}_{x}\sum_{l_j\in\mathcal{S}_{\nu}}\delta(y-l_{j})\frac{\partial^{2}}{\partial
x^{2}}P(x,y,t)+\mathcal{D}_{y}\mathcal{F}_{\kappa_{x}}^{-1}\left[|\kappa_{x}|^{1-\bar{\nu}}\frac{\partial^{2}}{\partial y^{2}}P(\kappa_x,y,t)\right].
\end{align}
Applying the Laplace and Fourier transforms, one obtains  Eq.~(\ref{fractal grid comb}) in the $(\kappa_x,y,s)$ space
\begin{align}\label{diffusion like eq on a comb Fourier-Laplace transform}
sP(\kappa_x,y,s)-\delta(y)
=-\mathcal{D}_{x}\kappa_{x}^{2}\sum_{l_j\in\mathcal{S}_{\nu}}\delta(y-l_j)P(\kappa_x,y,s)+\mathcal{D}_{y}|\kappa_{x}|^{1-\bar{\nu}}\frac{\partial^{2}}{\partial y^{2}}P(\kappa_x,y,s),
\end{align}
where $P(\kappa_{x},y,t=0)=\delta(y)$. By integrating over the $y$ coordinate, we analyze the reduced PDF $p_{1}(x,t)$ for the backbone dynamics. From Eq.~(\ref{diffusion like eq on a comb Fourier-Laplace transform}) we find
\begin{align}\label{p1 general}
p_{1}(\kappa_{x},s)=\frac{1}{s}\left[1-\mathcal{D}_{x}\kappa_{x}^{2}\sum_{l_j\in\mathcal{S}_{\nu}}P(\kappa_{x},y=l_{j},s)\right].
\end{align}
Presenting PDF image $P(\kappa_x,y,s)$ in the form of the ansatz
\begin{align}
P(\kappa_x,y,s)=g(\kappa_x,s)\exp\left(-\sqrt{\frac{s}{\mathcal{D}_{y}|\kappa_x|^{1-\bar{\nu}}}}|y|\right),
\end{align}
one obtains for a single backbone
\begin{align}P(\kappa_x,y=l_{j},s)=g(\kappa_x,s)\exp\left(-\sqrt{\frac{s}{\mathcal{D}_{y}|\kappa_x|^{1-\bar{\nu}}}}|l_{j}|\right).
\end{align}
Therefore, reduced PDF $p_{1}(x,s)$ reads
\begin{align}\label{p1 new}
p_{1}(\kappa_x,s)=\int_{-\infty}^{\infty}dy\,P(\kappa_x,y,s)=2g(\kappa_x,s)\sqrt{\frac{\mathcal{D}_{y}|\kappa_x|^{1-\bar{\nu}}}{s}}.
\end{align}

The summation in Eq.~(\ref{p1 general}) is over the fractal set, which corresponds to integration over the fractal measure $\mu_{\nu}\sim l^{\nu}$ ($d\mu_{\nu}=\frac{1}{\Gamma(\nu)}l^{\nu-1}dl$). This yields
\begin{align}\label{summation p lj}
\sum_{l_{j}\in\mathcal{S}_{\nu}}P(\kappa_x,y=l_{j},s)&=g(\kappa_x,s)\frac{1}{\Gamma(\nu)}\int_{0}^{\infty}dl\,l^{\nu-1}\exp\left(-\sqrt{\frac{s}{\mathcal{D}_{y}|\kappa_{x}|^{1-\bar{\nu}}}}l\right)\nonumber\\&=g(\kappa_x,s)\left(\frac{\mathcal{D}_{y}|\kappa_{x}|^{1-\bar{\nu}}}{s}\right)^{\nu/2}=\frac{1}{2\mathcal{D}_{y}^{\frac{1-\nu}{2}}}\frac{s^{\frac{1-\nu}{2}}}{|\kappa_{x}|^{(1-\nu)(1-\bar{\nu})/2}}p_{1}(\kappa_x,s).
\end{align}
Substituting relation (\ref{summation p lj}) in Eq.~(\ref{p1 general}), we obtain
\begin{align}\label{p1 fractal Laplace final}
s^{\frac{-1+\nu}{2}}\left[sp_{1}(\kappa_x,s)-p_{1}(\kappa_{x},t=0)\right]=-\frac{\mathcal{D}_{x}}{2\mathcal{D}_{y}^{\frac{1-\nu}{2}}}|\kappa_{x}|^{\alpha}p_{1}(\kappa_{x},s),
\end{align}
where $\alpha=\frac{4-(1-\nu)(1-\bar{\nu})}{2}$. 

The inverse Fourier-Laplace transforms yields the generalized diffusion equation
\begin{align}\label{p1 fractal final}
\int_{0}^{t}dt'\,\zeta(t-t')\frac{\partial}{\partial t'}p_{1}(x,t')=\frac{\mathcal{D}_{x}}{2\mathcal{D}_{y}^{\frac{1-\nu}{2}}}\frac{\partial^{\alpha}}{\partial|x|^{\alpha}}p_{1}(x,t),\nonumber\\
\end{align}
with the space fractional Riesz derivative $\frac{\partial^{\alpha}}{\partial|x|^{\alpha}}$\footnote{The Riesz fractional derivative of order $\alpha$ ($0<\alpha\leq2$) is given as a pseudo-differential operator with the Fourier symbol
$-|\kappa|^\alpha$, $\kappa\in \mathrm{R}$, i.e., $\frac{\partial^\alpha}{\partial|x|^\alpha}f(x)=\mathcal{F}^{-1}\left[-|\kappa|^\alpha f(\kappa)\right]$ \cite{SKM book}.}, and the convolution kernel
\begin{align}
\zeta(t)=\mathcal{L}^{-1}\left[s^{\frac{-1+\nu}{2}}\right]=\frac{t^{-\mu}}{\Gamma\left(1-\mu\right)},
\end{align}
where $\mu=(1+\nu)/2$.

Eventually, we obtain that the PDF $p_{1}(x,t)$ satisfies the following space-time fractional diffusion equation 
\begin{equation}\label{p1 fractal final2}
\frac{\partial^{\mu}}{\partial
t^{\mu}}p_{1}(x,t)=\frac{\mathcal{D}_{x}}{2\mathcal{D}_{y}^{\frac{1-\nu}{2}}}\frac{\partial^{\alpha}}{\partial|x|^{\alpha}}p_{1}(x,t),
\end{equation}
where $\frac{\partial^{\mu}}{\partial
t^{\mu}}$ is the Caputo time fractional derivative of
order $\mu=(1+\nu)/2$, where $1/2<\mu<1$\footnote{The Caputo fractional derivative of order $0<\mu<1$ is defined by
$\frac{\partial^{\mu}}{\partial
t^{\mu}}f(t)=\frac{1}{\Gamma(1-\mu)}\int_{0}^{t}d\tau\,(t-\tau)^{-\mu}
\frac{d}{d\tau}f(\tau)$
\cite{Caputo}.}. The solution of Eq.~(\ref{p1 fractal final2}) can be found in the form
of the Fox $H$-function $H_{p,q}^{m,n}(z)$ \cite{saxena book} (see Appendix \ref{app solution}). Therefore, the reduced PDF $p_1(x,t)$ reads
\begin{align}\label{PDF infinite backbones fractal comb}
p_{1}(x,t)=\mathcal{F}^{-1}\left[E_{\mu}\left(-\mathcal{D}_{\mu,\alpha}t^{\mu}|\kappa_{x}|^{\alpha}\right)\right]=\frac{1}{\alpha|x|} H_{3,3}^{2,1}\left[\left.\frac{|x|}{\left(\mathcal{D}_{\mu,\alpha}t^{\mu}\right)^{\frac{1}{\alpha}}}\right|\begin{array}{c l}
    (1,\frac{1}{\alpha}),(1,\frac{1+\nu}{2\alpha}),(1,\frac{1}{2})\\
    (1,1),(1,\frac{1}{\alpha}),(1,\frac{1}{2})
  \end{array}\right],
\end{align}
where
$\mathcal{D}_{\mu,\alpha}=\mathcal{D}_{x}/2\mathcal{D}_{y}^{1-\mu}$
is the generalized diffusion coefficient with physical dimension
$[\mathcal{D}_{\nu}]=\mathrm{m}^{\alpha}/\mathrm{s}^{\mu}$, and $E_{\alpha}(z)=E_{\alpha,1}(z)$ is the one parameter Mittag-Leffler function \cite{erdelyi}. 

Since the second moment does not exist, we can calculate the fractional moment $\left\langle |x|^{q}\right\rangle=\int_{-\infty}^{\infty}dx\,|x|^{q}p_{1}(x,t)$, $0<q<\alpha<2$, and then as the MSD we calculate the fractional moment $\left\langle |x|^{q}\right\rangle^{2/q}$. Thus, it is obtained (see Appendix \ref{app solution})
\begin{align}\label{fractional moments fractal comb}
\left\langle |x|^{q}\right\rangle^{\frac{2}{q}}=\left[\frac{2}{\alpha}\left(\mathcal{D}_{\mu,\alpha}t^{\mu}\right)^{\frac{q}{\alpha}}\frac{\Gamma(1+q)\sin\left(\frac{q\pi}{2}\right)}{\Gamma\left(1+\frac{\mu q}{\alpha}\right)\sin\left(\frac{q\pi}{\alpha}\right)}\right]^{\frac{2}{q}}\sim t^{\frac{2\mu}{\alpha}},
\end{align}
where $\frac{2\mu}{\alpha}=\frac{2(1+\nu)}{4-(1-\nu)(1-\bar{\nu})}$. Therefore, the fractal distribution of the backbones changes the transport exponent. When $\bar{\nu}=1$, which corresponds to the continuous distribution of fingers and $\alpha=2$, one arrives at the result of Ref.~\cite{fractal grid pre} for the fractal grid comb $\left\langle x^{2}(t)\right\rangle\sim t^{\frac{1+\nu}{2}}$. For $\nu=0$ (one backbone), one finds $\left\langle |x|^{q}\right\rangle^{2/q}\sim t^{\frac{2}{3+\bar{\nu}}}$ \cite{jpa2016}. We conclude here that the fractal structures of both backbones and the fingers increase the transport exponent of subdiffusion along the fractal backbone structure.

\section{Superdiffusion due to compensation kernel}\label{superdiffusion}

\begin{figure}\resizebox{0.5\textwidth}{!}{\includegraphics{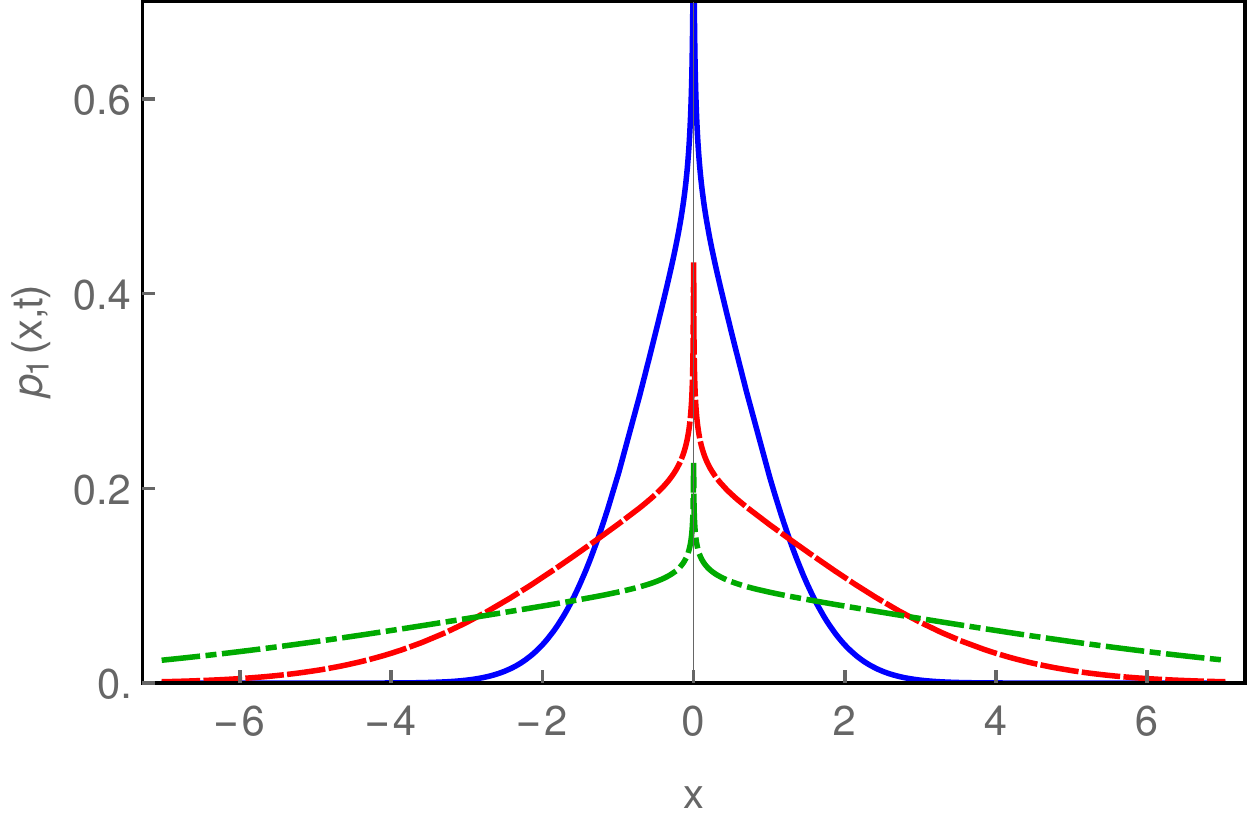}} \caption {Graphical representation of the PDF (\ref{p1(x,t) final eta}) for $\nu=\bar{\nu}=1/2$, $\mathcal{D}_{x}=1$, $\mathcal{D}_{y}=1$, and $t=1$ (solid blue line), $t=5$ (red dashed line), $t=20$ (green dot-dashed line).}\label{pdf mesh eta}
\end{figure}

In what follows we consider fractal mesh model with compensation memory kernel $\eta(t)$. As it has been shown in Ref.~\cite{jpa2016} the general compensation kernel with combination with the fractal structure of fingers may be responsible for superdiffusion in the system. In this case, the kernel compensates the trapping of particle in the fingers, for example as a result of a complex environment which accelerates the contaminant \cite{jpa2016}. In particular, the random walks in comb structure can be used in modeling of the RNA polymerase transcription, where the $x$-axis of backbones and the $y$-axis of fingers correspond to the active transcription and back-tracking, respectively \cite{jaeoh}. Therefore, we introduce the memory kernel inside the backbone dynamics  of the fractal mesh model (\ref{fractal mesh})
{\small{\begin{align}\label{fractal mesh eta} 
\frac{\partial}{\partial t}P(x,y,t)
=\mathcal{D}_{x}\sum_{l_j\in\mathcal{S}_{\nu}}\delta(y-l_{j})\int_{0}^{t}dt'\eta(t-t')\frac{\partial^{2}}{\partial
x^{2}}P(x,y,t')+\mathcal{D}_{y}\sum_{r_k\in\mathcal{S}_{\nu}}\delta(x-r_{k})\frac{\partial^{2}}{\partial y^{2}}P(x,y,t).
\end{align}}}Employing the Laplace transform of Eq.~(\ref{fractal mesh eta}), and using the separation ansatz (\ref{ansatz}), and integration of the equation over $y$, we find that the reduced PDF $p_{1}(x,t)$ is governed by the following equation in the Laplace space
\begin{align}\label{fractal mesh L f eta} 
s|x|^{-\bar{\nu}(1-\nu)/2}f(x,s)-\frac{\mathcal{D}_{x}}{2\left(\mathcal{D}_{y}C_{\bar{\nu}}\right)^{(1-\nu)/2}}\eta(s)s^{(1-\nu)/2}\frac{\partial^{2}}{\partial
x^{2}}f(x,s)=\delta(x),
\end{align}
where $f(x,s)=|x|^{\bar{\nu}(1-\nu)/2}p_{1}(x,s)$. Let us use the compensate kernel in the form 
\begin{align}\label{eta}
\eta(t)=t^{-\mu}/\Gamma(1-\mu), \quad \mu=(1+\nu)/2,
\end{align}
which also accounts the fractal structure of the backbones reflected in the transport exponent $\mu$. Therefore, the solution for the reduced PDF reads
\begin{align}\label{p1(x,t) final eta}
p_{1}(x,t)=\frac{\left(\frac{2\left(\mathcal{D}_{y}C_{\bar{\nu}}\right)^{1-\mu}}{\mathcal{D}_{x}}\right)^{1-\frac{1}{\alpha}}}{2\alpha^{1-2/\alpha}\Gamma\left(1-1/\alpha\right)}\frac{|x|^{\alpha-2}}{t^{1-\frac{1}{\alpha}}}\exp\left(-\frac{2\left(\mathcal{D}_{y}C_{\bar{\nu}}\right)^{1-\mu}}{\alpha^{2}\mathcal{D}_{x}}\frac{|x|^{\alpha}}{t}\right).
\end{align}
Details of the inferring of the solution (\ref{p1(x,t) final eta}) is presented in Appendix~\ref{app 3}. Graphical representation of the PDF (\ref{p1(x,t) final eta}) is given in Fig.~\ref{pdf mesh eta}.

From the reduced PDF (\ref{p1(x,t) final eta}), we calculate the MSD, which results to superdiffusion
\begin{align}\label{msd w eta}
\left\langle x^{2}(t)\right\rangle\simeq t^{\frac{4}{4-\bar{\nu}(1-\nu)}},
\end{align}
with transport exponent $1<\frac{4}{4-\bar{\nu}(1-\nu)}$ for any $0<\nu<1$ and $0<\bar{\nu}<1$. For $\bar{\nu}=0$ (continuous distribution of fingers), normal diffusion takes place for any $\nu$. This effect is due to the memory kernel which compensates the trapping effect of fingers. Obviously, that as the result of this compensation superdiffusion takes place when $\bar{\nu}\neq 0$. The case with $\nu=0$, which corresponds to the finite number of backbones, yields $\left\langle x^{2}\right\rangle\sim t^{\frac{4}{4-\bar{\nu}}}$ \cite{jpa2016}. Thus, the compensation kernel, and the fractal structure of both backbones and the fingers are responsible for appearance of superdiffusion.

In a similar way, the compensation kernel $\eta(t)$ can be considered for the fractal grid comb  (\ref{fractal grid comb}), which now reads
{\small{\begin{align}\label{fractal mesh2} 
\frac{\partial}{\partial t}P(x,y,t)
=\mathcal{D}_{x}\sum_{l_j\in\mathcal{S}_{\nu}}\delta(y-l_{j})\int_{0}^{t}dt'\,\eta(t-t')\frac{\partial^{2}}{\partial
x^{2}}P(x,y,t')+\mathcal{D}_{y}\mathcal{F}_{\kappa_{x}}^{-1}\left[|\kappa_{x}|^{1-\bar{\nu}}\frac{\partial^{2}}{\partial y^{2}}P(\kappa_x,y,t)\right].
\end{align}}}Let us consider $\eta(t)=t^{-(1+\nu)/2}/\Gamma\left(1-\frac{1}{2}-\frac{\nu}{2}\right)$. Following the same procedure of solving Eq.~(\ref{fractal grid comb}) we find, from (\ref{fractal mesh2}) that the PDF $p_{1}(x,t)$ is governed by the space fractional diffusion equation
\begin{equation}\label{p1 fractal final3}
\frac{\partial}{\partial
t}p_{1}(x,t)=\frac{\mathcal{D}_{x}}{2\mathcal{D}_{y}^{(1-\nu)/2}}\frac{\partial^{\alpha}}{\partial|x|^{\alpha}}p_{1}(x,t),
\end{equation}
and its solution is given in the form the Fox $H$-function (see Appendix \ref{app solution})
\begin{align}\label{PDF infinite backbones fractal comb2}
p_{1}(x,t)=\mathcal{F}^{-1}\left[\exp\left(-\frac{\mathcal{D}_{x}}{2\mathcal{D}_{y}^{(1-\nu)/2}}t|\kappa_{x}|^{\alpha}\right)\right]=\frac{1}{\alpha|x|} H_{3,3}^{2,1}\left[\left.\frac{|x|}{\left(\frac{\mathcal{D}_{x}}{2\mathcal{D}_{y}^{(1-\nu)/2}}t\right)^{1/\alpha}}\right|\begin{array}{c l}
    (1,\frac{1}{\alpha}),(1,\frac{1}{\alpha}),(1,\frac{1}{2})\\
    (1,1),(1,\frac{1}{\alpha}),(1,\frac{1}{2})
  \end{array}\right].
\end{align}
Thus, the $q$-th moment yields (see Appendix \ref{app solution})
\begin{align}\label{fractional moments fractal comb2}
\left\langle |x|^{q}\right\rangle^{\frac{2}{q}}=\left[\frac{2}{\alpha}\left(\frac{\mathcal{D}_{x}}{2\mathcal{D}_{y}^{(1-\nu)/2}}\right)^{q/\alpha}\frac{t^{q/\alpha}}{\Gamma(1+q/\alpha)}\right]^{\frac{2}{q}}\sim t^{\frac{4}{4-(1-\nu)(1-\bar{\nu})}}.
\end{align}
Therefore, superdiffusion takes place for any $0<\nu<1$ and $0<\bar{\nu}<1$. This is a typical result for the L\'{e}vy distribution. For $\bar{\nu}=1$ (continuous distribution of fingers), the normal diffusive behavior is obtained for any $\nu$, which results from the compensation kernel effect. For a finite number of backbones, one obtains $\left\langle |x|^{q}\right\rangle^{2/q}\sim t^{\frac{4}{3+\bar{\nu}}}$. Therefore, the compensation kernel, and the fractal structure of both backbones and fingers are responsible for superdiffusion.

\section{Summary}

In this research we concern with anomalous diffusion on a fractal mesh with inhomogeneous (fractal) distribution of both fingers and backbones. A specific property of this fractal mesh is that both fingers and backbones are distributed (with the power law scaling) along entire $x$ and $y$ axes from $-\infty$ to $+\infty$. However, the main transport channels are backbones, while fingers play a role of traps. Therefore, we have studied the transport properties, namely the transport exponents of the system as a function of the fractal dimensions of the backbones $\nu$ and fingers $\bar{\nu}$. We presented exact results for the probability distribution function and the mean square displacement in the case when the fractal mesh structure of the fingers is controlled by the Weierstrass function. Additionally, we considered a fractal grid model with the power-law distribution of the fingers. Exact expressions for the probability distribution functions, which describe fractional kinetics in the inhomogeneous media, is obtained, and the mean squared displacement, and the q-th moments are derived as the main characteristics of the particle behavior. Our main result is that for all possible realizations of the fractal mesh, or grid structures, the transport exponent $\beta$  is unambiguously determined by the fractal dimensions of the fingers and backbones $\bar{\nu}$ and $\nu$ in Eqs.~(\ref{msd w}) and (\ref{fractional moments fractal comb}). Another important result relates with superdiffusion in fractal mesh due to a compensation kernel. The compensation kernel can appear in the system as a result of complex environment, which accelerates the contaminant spreading in the inhomogeneous media. The interplay between the compensation kernel and the fractal structure of the backbones and fingers leads to superdiffusion. We obtain analytical expression for the transport exponent, which is determined by the fractal dimensions of the fractal mesh structure and an analytical expression for the  PDF of the contaminant spreading is obtained in terms of the Fox $H$-functions.

\section*{Acknowledgment}

TS and AI thank the hospitality at the Max-Planck Institute for the Physics of Complex Systems in Dresden, Germany where this work has been done. AI was also supported by the Israel Science Foundation (ISF).

\appendix


\section{Solution of space-time fractional diffusion equation. Fox $H$-function}\label{app solution}

Let us consider the following space-time fractional diffusion equation
\begin{eqnarray}\label{Eq.100}
\frac{\partial^{\lambda}}{\partial t^{\lambda}}
W(x,t)=\mathcal{D}_{\lambda,\alpha}\frac{\partial^\alpha}{\partial
|x|^\alpha}W(x,t), \quad t>0, \quad -\infty<x<+\infty,
\end{eqnarray}
where $\frac{\partial^{\lambda}}{\partial t^{\lambda}}$ is Caputo time fractional derivative of order $0<\lambda<1$, $\frac{\partial^\alpha}{\partial
|x|^\alpha}$ is the Riesz space fractional derivative of order $1<\alpha<2$, and $\mathcal{D}_{\lambda,\alpha}$ is the generalized diffusion coefficient with physical dimension $\left[\mathcal{D}_{\lambda,\alpha}\right]=\mathrm{m}^{\alpha}\mathrm{s}^{-\lambda}$. The boundary conditions at infinities are
\begin{eqnarray}\label{Eq.200}
W(\pm\infty,t)=0, \quad \frac{\partial}{\partial x}W(\pm\infty,t)=0, \quad t>0,
\end{eqnarray}
and the initial condition is
\begin{eqnarray}\label{Eq.300}
W(x,0)=\delta(x), \quad -\infty<x<+\infty.
\end{eqnarray}
Applying the Fourier-Laplace transform in Eq.~(\ref{Eq.100}), one finds
\begin{eqnarray}\label{laplacefourier100}
W(\kappa,s)=\frac{s^{\lambda-1}}{s^{\lambda}+\mathcal{D}_{\lambda,\alpha}|\kappa|^\alpha}.
\end{eqnarray}
where we take into consideration the initial condition (\ref{Eq.300}) and the boundary conditions (\ref{Eq.200}). Note that here we use the property of the Laplace transform of the Caputo derivative \cite{Caputo}
\begin{eqnarray}\label{laplace Caputo}
\mathcal{L}\left[\frac{\partial^{\lambda}}{\partial t^{\lambda}}f(t)\right]=s^{\lambda}\mathcal{L}\left[f(t)\right]-s^{\lambda-1}f(0).
\end{eqnarray}
Accounting the definition of the Mittag-Leffler function by means of the Laplace inversion \cite{saxena book}
\begin{eqnarray}
\label{ML three Laplace}
\mathcal{L}^{-1}\left[\frac{s^{\alpha-1}}{s^{\alpha}\mp a}\right]=E_{\alpha}(\pm at^{\alpha}),
\end{eqnarray}
for $\Re(s)>|a|^{1/\alpha}$, where $E_{\alpha}(z)$ is the one parameter Mittag-Leffler function, one arrives at the solution in the form of the Fox $H$ function
\begin{eqnarray}\label{inverselaplace100}
W(\kappa,t)=E_{\lambda}\left(-\mathcal{D}_{\lambda,\alpha}t^{\lambda}|\kappa|^\alpha
\right)=H_{1,2}^{1,1}\left[\mathcal{D}_{\lambda,\alpha}t^{\lambda}|\kappa|^\alpha\left|
\begin{array}{l}(0,1)\\(0,1),(0,\lambda)\end{array}\right.\right].
\end{eqnarray}
Here $H_{p,q}^{m,n}(z)$ is Fox $H$-function, defined as the inverse Mellin transform for a set of gamma functions \cite{saxena book}
\begin{eqnarray}
H_{p,q}^{m,n}\left[z\left|\begin{array}{c l}
    (a_p,A_p)\\
    (b_q,B_q)
  \end{array}\right.\right]=H_{p,q}^{m,n}\left[z\left|\begin{array}{l}(a_1,A_1),\ldots,(a_p,A_p)\\
(b_1,B_1),\ldots,b_q,B_q)\end{array}\right.\right]=\frac{1}{2\pi\imath}\int_{\Omega}ds\,\theta(s)z^{-s},
\label{H_integral}
\end{eqnarray}
where
\begin{eqnarray}\label{theta}
\theta(s)=\frac{\prod_{j=1}^{m}\Gamma(b_j+B_js)\prod_{j=1}^{n}\Gamma(1-a_j-A_js)}{
\prod_{j=m+1}^{q}\Gamma(1-b_j-B_js)\prod_{j=n+1}^{p}\Gamma(a_j+A_js)},
\end{eqnarray}
with $0\leq n\leq p$, $1\leq m\leq q$, $a_i,b_j \in C$, $A_i,B_j\in R^{+}$, $i=1,
\ldots,p$, and $j=1,\ldots,q$. The contour $\Omega$ starting at
$c-i\infty$ and ending at $c+i\infty$ separates the poles
of the function $\Gamma(b_j+B_js)$, $j=1,...,m$ from those of the
function $\Gamma(1-a_i-A_is)$, $i=1,...,n$. The expansion for the
$H$-function (\ref{H_integral}) is given by \cite{saxena book}
\begin{align}\label{H_expansion}
H_{p,q}^{m,n}\left[z\left|\begin{array}{c l}
    (a_p,A_p)\\
    (b_q,B_q)
  \end{array}\right.\right]=\sum_{h=1}^{m}\sum_{k=0}^{\infty}\frac{\prod_{j=1, j\neq
h}^{m}\Gamma\left(b_j-B_j\frac{b_h+k}{B_h}\right)\prod_{j=1}^{n}\Gamma\left(1-a_j+A_j\frac{b_h+k}{B_h}\right)}{\prod_{j=m+1}^{q}\Gamma\left(1-b_j+B_j\frac{b_h+k}{B_h}\right)\prod_{j=n+1}^{p}\Gamma\left(a_j-A_j\frac{b_h+k}{B_h}\right)}\cdot\frac{(-1)^kz^{(b_h+k)/B_h}}{k!\,B_h}.
\end{align}

From the inverse Fourier transform, by using the Mellin-cosine transform of Fox $H$-function \cite{saxena book}
\begin{eqnarray}
\label{cosine H}
\int_{0}^{\infty}{d}\kappa\,\kappa^{\rho-1}\cos(\kappa x)H_{p,q}^{m,n}\left[a\kappa^{\delta}\left|
\begin{array}{l}(a_p,A_p)\\(b_q,B_q)\end{array}\right.\right]=\frac{\pi}{x^\rho}H_{q+1,p+2}^{n+1,m}\left[\frac{x^\delta}{a}\left|
\begin{array}{l}(1-b_q,B_q),(\frac{1+\rho}{2},\frac{\delta}{2})\\(\rho,\delta),
(1-a_p,A_p),(\frac{1+\rho}{2},\frac{\delta}{2})\end{array}\right.\right],
\end{eqnarray}
one obtains
\begin{eqnarray}\label{special case}
W(x,t)=\frac{1}{\alpha
|x|}H_{3,3}^{2,1}\left[\frac{|x|}{\left(\mathcal{D}_{\lambda,\alpha}t^{\lambda}\right)^{1/\alpha}}
\left|\begin{array}{l}
    (1,\frac{1}{\alpha}),(1,\frac{\lambda}{\alpha}),(1,\frac{1}{2})\\
    (1,1),(1,\frac{1}{\alpha}),(1,\frac{1}{2})
  \end{array}\right.\right].
\end{eqnarray}

From solution (\ref{special case}) the fractional $q$ moments can be calculated. Performing the Mellin transform of the Fox $H$-function
\begin{eqnarray}
\int_0^{\infty}dx\,x^{\xi-1}H_{p,q}^{m,n}\left[ax\left|\begin{array}{l}(a_p,A_p)\\
(b_q,B_q)\end{array}\right.\right]=a^{-\xi}\theta(\xi),
\label{integral of H}
\end{eqnarray}
where $\theta(\xi)$ is defined in Eqs.~(\ref{H_integral}) and (\ref{theta}), one obtains for the fractional moments
\begin{align}\label{moments case}
\left\langle |x|^{q}(t)\right\rangle&=\frac{2}{\alpha}\int_{0}^{\infty}dx\,x^{q-1}H_{3,3}^{2,1}\left[\frac{x}{\left(\mathcal{D}_{\lambda,\alpha}t^{\lambda}\right)^{1/\alpha}}
\left|\begin{array}{l}
    (1,\frac{1}{\alpha}),(1,\frac{\lambda}{\alpha}),(1,\frac{1}{2})\\
    (1,1),(1,\frac{1}{\alpha}),(1,\frac{1}{2})
  \end{array}\right.\right]\nonumber\\ &=\frac{2}{\alpha}\left(\mathcal{D}_{\lambda,\alpha}t^{\lambda}\right)^{q/\alpha}\theta(q)
=\frac{4}{\alpha}\cdot\frac{\Gamma\left(q\right)\Gamma(1+q/\alpha)\Gamma(-q/\alpha)}{\Gamma(q/2)\Gamma(-q/2)}\cdot
\frac{\left(\mathcal{D}_{\lambda,\alpha}t^\lambda\right)^{q/\alpha}}{\Gamma\left(1+\frac{\lambda q}{\alpha}\right)},\nonumber\\
\end{align}
where
\begin{eqnarray}\label{theta ex}
\theta(q)=\frac{\Gamma(1+q)\Gamma(1+q/\alpha)\Gamma(-q/\alpha)}{\Gamma(-q/2)\Gamma(1+\lambda q/\alpha)\Gamma(1+q/2)} =\frac{2\Gamma(q)\Gamma(1+q/\alpha)\Gamma(-q/\alpha)}{\Gamma(-q/2)\Gamma(1+\lambda q/\alpha)\Gamma(q/2)}.
\end{eqnarray}

\section{Solution of the Lommel equation in terms of the Fox functions}\label{app 2}

The solution of the Lommel differential equation 
\begin{align}\label{Lommel}
u''(x)-c^{2}x^{2\zeta-2}u(x)=0
\end{align}
is given in terms of the Bessel functions \cite{book integrals}
\begin{align}\label{modified Bessel}
u(x)=\sqrt{x}Z_{\frac{1}{2\zeta}}\left(\imath\frac{c}{\zeta}x^{\zeta}\right).
\end{align}
The Bessel function $Z_{\frac{1}{2\zeta}}(x)$ is given by $Z_{\frac{1}{2\zeta}}(x)=C_{1}J_{\frac{1}{2\zeta}}(x)+C_{2}N_{\frac{1}{2\zeta}}(x)$, where $J_{\frac{1}{2\zeta}}(x)$ is the Bessel function of the first kind and $N_{\frac{1}{2\zeta}}(x)$ is the Bessel function of the second kind (Neumann function). In our case, the Bessel function is with imaginary argument, therefore the solution of the Lommel equation (\ref{Lommel}) is given in terms of the modified Bessel function (of the third kind) \cite{book integrals}
\begin{align}\label{Bessel third kind}
u(x)=\sqrt{x}K_{\frac{1}{2\zeta}}\left(\frac{c}{\zeta}x^{\zeta}\right),
\end{align}
which satisfies the zero boundary conditions at infinity. 

The modified Bessel function (of the third kind) $K_{\nu}(z)$ is a special case of the Fox $H$-function \cite{saxena book}
\begin{eqnarray}
H_{0,2}^{2,0}\left[\frac{z^{2}}{4}\left|\begin{array}{l} \\
(\frac{a+\nu}{2},1),(\frac{a-\nu}{2},1)\end{array}\right.\right]=2\left(\frac{z}{2}\right)^{a}K_{\nu}(z).
\label{HK relation}
\end{eqnarray}
Its series representation is given by
\begin{eqnarray}
K_{\nu}(z)\simeq\frac{\Gamma(\nu)}{2}\left(\frac{z}{2}\right)^{-\nu}\left[1+\frac{z^{2}}{4(1-\nu)}+\dots\right]+\frac{\Gamma(-\nu)}{2}\left(\frac{z}{2}\right)^{\nu}\left[1+\frac{z^{2}}{4(\nu+1)}+\dots\right], \quad z\rightarrow0, \quad \nu\notin Z.
\label{K series}
\end{eqnarray}

The inverse Laplace transform of the Fox $H$-function is used in order to find solution (\ref{p1(x,t) final exact}) from Eq.~(\ref{f(x,s) green sol K delta}). The Laplace transform reads \cite{saxena book}
\begin{eqnarray}
\mathcal{L}^{-1}\left[s^{-\rho}H_{p,q}^{m,n}\left[as^{\sigma}\left|\begin{array}{l}(a_p,A_p)\\
(b_q,B_q)\end{array}\right.\right]\right]=t^{\rho-1}H_{p+1,q}^{m,n}\left[\frac{a}{t^{\sigma}}\left|\begin{array}{l}(a_p,A_p),(\rho,\sigma)\\
(b_q,B_q)\end{array}\right.\right],
\label{H_laplace}
\end{eqnarray}

\section{Solution of Eq.~(\ref{fractal mesh L f eta})}\label{app 3}

Presenting details of the solution of Eq.~(\ref{fractal mesh L f eta}), we first consider the homogeneous equation 
\begin{align}\label{fractal mesh L f eta homogeneous} 
s|x|^{-\bar{\nu}(1-\nu)/2}G(x,s)-\frac{\mathcal{D}_{x}}{2\left(\mathcal{D}_{y}C_{\bar{\nu}}\right)^{(1-\nu)/2}}s^{\mu-1}s^{(1-\nu)/2}\frac{\partial^{2}}{\partial
x^{2}}G(x,s)=0,
\end{align}
where we use $\eta(t)=t^{-\mu}/\Gamma(1-\mu), \quad \mu=(1+\nu)/2$. This is a Lommel equation (\ref{Lommel}). Its solution is given in terms of the modified Bessel function or Fox $H$-function
\begin{align}\label{f(x,s) green sol K eta}
G(x,s)=\sqrt{x}K_{\frac{1}{\alpha}}\left(\frac{2s^{1/2}x^{\alpha/2}}{\alpha}\sqrt{\frac{2\left(\mathcal{D}_{y}C_{\bar{\nu}}\right)^{1-\mu}}{\mathcal{D}_{x}}}\right)=\frac{\sqrt{x}}{2}H_{0,2}^{2,0}\left[\left.\frac{s\,x^{\alpha}}{\alpha^{2}}\frac{2\left(\mathcal{D}_{y}C_{\bar{\nu}}\right)^{1-\mu}}{\mathcal{D}_{x}}\right|\left.\begin{array}{l} \\
(\frac{1}{2\alpha},1),(-\frac{1}{2\alpha},1)\end{array}\right.\right],
\end{align} 
where $\alpha=\frac{4-\bar{\nu}(1-\nu)}{2}$, $\mu=(1+\nu)/2$. In order to solve the inhomogeneous Lommel Eq.~(\ref{fractal mesh L f eta}), we use the function $f(|x|,s)=\mathcal{C}(s)G(|x|,s)=\mathcal{C}(s)G(y,s)$, where $G(y,s)$ is given by (\ref{f(x,s) green sol K eta}), $y=|x|$, and $\mathcal{C}(s)$ is a function which depends on $s$. Therefore, we find 
\begin{align}\label{condition}
-2\left[\frac{\mathcal{D}_{x}}{2\left(\mathcal{D}_{y}C_{\bar{\nu}}\right)^{1-\mu}}\right]\frac{\partial}{\partial y}f(y=0,s)=1.
\end{align}
From the series representation of the modified Bessel function (\ref{K series}), we find $\mathcal{C}(s)$. By inverse Laplace transform (see relation (\ref{H_laplace}) in Appendix \ref{app 2}) we arrive at the solution for the reduced PDF
\begin{align}\label{p1(x,t) final eta 2}
p_{1}(x,t)=\frac{\left(\frac{2\left(\mathcal{D}_{y}C_{\bar{\nu}}\right)^{1-\mu}}{\mathcal{D}_{x}}\right)^{1-\frac{1}{2\alpha}}}{2\alpha^{1-1/\alpha}\Gamma\left(1-1/\alpha\right)}\frac{|x|^{\alpha-3/2}}{t^{1-\frac{1}{2\alpha}}} H_{0,1}^{1,0}\left[\left.\frac{1}{\alpha^{2}}\frac{2\left(\mathcal{D}_{y}C_{\bar{\nu}}\right)^{1-\mu}}{\mathcal{D}_{x}}\frac{x^{\alpha}}{t}\right|\left.\begin{array}{l} \\
(-\frac{1}{2\alpha},1)\end{array}\right.\right].
\end{align}
Using the property of the Fox $H$-function \cite{saxena book}
\begin{align}\label{H_property}
H_{p,q}^{m,n}\left[z^{\delta}\left|\begin{array}{c l}
    (a_p,A_p)\\
    (b_q,B_q)
  \end{array}\right.\right]=\frac{1}{\delta}\cdot H_{p,q}^{m,n}\left[z\left|\begin{array}{c l}
    (a_p,A_p/\delta)\\
    (b_q,B_q/\delta)
  \end{array}\right.\right],
\end{align}
we rewrite solution (\ref{p1(x,t) final eta 2}) as
\begin{align}\label{p1(x,t) final eta 22}
p_{1}(x,t)=\frac{\left(\frac{2\left(\mathcal{D}_{y}C_{\bar{\nu}}\right)^{1-\mu}}{\mathcal{D}_{x}}\right)^{1-\frac{1}{2\alpha}}}{2\alpha^{2-1/\alpha}\Gamma\left(1-1/\alpha\right)}\frac{|x|^{\alpha-3/2}}{t^{1-\frac{1}{2\alpha}}} H_{0,1}^{1,0}\left[\left.\frac{1}{\alpha^{2/\alpha}}\left(\frac{2\left(\mathcal{D}_{y}C_{\bar{\nu}}\right)^{1-\mu}}{\mathcal{D}_{x}}\right)^{1/\alpha}\frac{|x|}{t^{1/\alpha}}\right|\left.\begin{array}{l} \\
(-\frac{1}{2\alpha},\frac{1}{\alpha})\end{array}\right.\right].
\end{align}
Then using another property of the Fox $H$-function \cite{saxena book}
\begin{align}\label{H_property2}
z^{\sigma}H_{p,q}^{m,n}\left[z\left|\begin{array}{c l}
    (a_p,A_p)\\
    (b_q,B_q)
  \end{array}\right.\right]=H_{p,q}^{m,n}\left[z\left|\begin{array}{c l}
    (a_p+\sigma A_p,A_p)\\
    (b_q+\sigma B_q,B_q)
  \end{array}\right.\right],
\end{align}
for $\sigma=1/2$, and again applying relation (\ref{H_property}), we rewrite solution (\ref{p1(x,t) final eta 22}) as follows
\begin{align}\label{p1(x,t) final eta h exp}
p_{1}(x,t)=\frac{\left(\frac{2\left(\mathcal{D}_{y}C_{\bar{\nu}}\right)^{1-\mu}}{\mathcal{D}_{x}}\right)^{1-\frac{1}{\alpha}}}{2\alpha^{1-2/\alpha}\Gamma\left(1-1/\alpha\right)}\frac{|x|^{\alpha-2}}{t^{1-\frac{1}{\alpha}}}
H_{0,1}^{1,0}\left[\left.-\frac{2\left(\mathcal{D}_{y}C_{\bar{\nu}}\right)^{1-\mu}}{\alpha^{2}\mathcal{D}_{x}}\frac{|x|^{\alpha}}{t}\right|\left.\begin{array}{l} \\
(0,1)\end{array}\right.\right].
\end{align}
Since the relation between the Fox $H$-function and the exponential function is given by
\begin{align}\label{H_exp}
H_{0,1}^{1,0}\left[z\left|\begin{array}{c l}
    \\
    (0,1)
  \end{array}\right.\right]=e^{-z},
\end{align} 
from Eq.~(\ref{p1(x,t) final eta h exp}) we finally obtain the solution (\ref{p1(x,t) final eta}).

\end{document}